\documentclass[a4paper]{jpconf}

\usepackage[ansinew]{inputenc}
\usepackage{amsfonts}
\usepackage{lmodern}
\usepackage{mathrsfs}
\usepackage[all]{xy}
\usepackage{amsthm, amssymb, amsmath, latexsym}
\usepackage{color}
\usepackage{array, float}
\usepackage{bm}
\usepackage{hyperref}
\usepackage{graphicx}
 \graphicspath{{Images/}}

\newtheorem{Theorem}{Theorem}[section]

\newtheorem{Lemma}[Theorem]{Lemma}

\theoremstyle{definition}
\newtheorem{Definition}[Theorem]{Definition}

\theoremstyle{remark}
\newtheorem{rem}[Theorem]{Remark}

\numberwithin{equation}{section}

\begin{document}

\setlength\parindent{0pt}


\title{Green operators for low regularity spacetimes}

\author{Yafet Sanchez Sanchez and James Vickers}
\address{Mathematical Sciences and STAG Research Centre, University of Southampton, Southampton, SO17 1BJ.}

\ead{Y.SanchezSanchez@soton.ac.uk and J.A.Vickers@soton.ac.uk}



\begin{abstract}
In this paper we define and construct advanced and retarded Green operators for the wave operator
 on spacetimes with  low regularity. In order to do so we require that the spacetime satisfies the condition 
of generalised hyperbolicity which is equivalent to well-posedness of the classical inhomogeneous problem
 with zero initial data where weak solutions are properly supported.
 Moreover, we provide an explicit formula for the kernel of the Green operators in terms
of an arbitrary eigenbasis of $H^1$ and a suitable Green matrix that solves a system of second order ODEs.
\end{abstract}


\section{Introduction}
The existence of Green operators for normally hyperbolic operators is
well understood for globally hyperbolic spacetimes where the spacetime
metric is a smooth \cite{bar}. However, there are two
important motivations for analysing spacetime metrics with finite
differentiability. Firstly, there are several models of physical
phenomena that require finite metric regularity. These include
impulsive gravitational waves, stars with well-defined surfaces,
general relativistic fluids and cosmic strings. Secondly, Einstein's
equations, viewed as a system of hyperbolic PDEs, can be naturally
formulated in function spaces with finite regularity.  

When the spacetime is of finite regularity the existence of Green
operators for the Klein-Gordon equation has been explored using
semigroup techniques in the Hamiltonian formalism \cite{fp}.  We
complement such an analysis by providing a definition of Green operators
subject only to the well posedness of the classical
problem. Furthermore we show how to explicitly construct both the
advanced and retarded the Green operators needed in order to formulate
quantum field theory on a non-smooth background.

In the first section of the paper we give the definition of the
advanced and retarded Green operators in the smooth case and define
the concept of generalised hyperbolicity. In the second section of the
paper we extend the definitions to the non-smooth setting and prove the
existence of the Green operators for low regularity spacetimes. In the
third section of the paper we give an explicit formula for the
advanced Green operator in the non-smooth setting.


\section{Preliminaries} \label{pre}


\subsection{The general setting}

We will consider a manifold $M$ of the form $M=(0,T)\times \Sigma$ and its closure $\overline{M}=[0,T]\times \Sigma$
where $\Sigma$ is a closed compact manifold. Given a smooth Lorentzian metric $g_{ab}$
of the form
\begin{equation}
ds^{2}=N^{2}dt^2 - {\gamma}_{ij} (dx^i +\beta^i dt)(dx^j +\beta^j dt)
\end{equation}
the wave operator $\square_g$ acting on a scalar
function $u$ is given by
\begin{equation}\label{square}
\begin{split}
  \square_{g}u=\frac{1}{N{\sqrt{\gamma}}}\left(\partial_{t}\left(N{\sqrt{\gamma}}\frac{1}{N^{2}}\partial_{t}u\right)\right)\\
  +\frac{1}{N{\sqrt{\gamma}}}\left(\partial_{t}\left(N{\sqrt{\gamma}}\frac{\beta^{i}}{N^{2}}\partial_{i}u\right)+\partial_{j}\left(N{\sqrt{\gamma}}\frac{\beta^{j}}{N^{2}}\partial_{t}u\right)\right)\\
  -\frac{1}{N{\sqrt{\gamma}}}\partial_{i}\left(N{\sqrt{\gamma}}({\gamma}^{ij}-\frac{\beta^{i}\beta^{j}}{N^{2}})\partial_{j}u\right)
 \end{split}
\end{equation}
where $N$ is the lapse function, $\beta^{i}$ the shift,
${\gamma}_{ij}$ the metric on $\Sigma$ and $\sqrt{\gamma} dx$ the
induced volume form on $\Sigma$. We use $a,b,c,d$ etc. to denote
spacetime indices and $i,j, k$ etc. to denote purely spatial indices.

The advanced zero initial data inhomogeneous problem for the wave
equation on ${\overline{M}}$  is given by

\begin{align}\label{wave}
  \square_{g}u&=f \mbox{ on $M$}\\\label{wave1}
 u(0,x)&=0 \mbox{ on $\Sigma_{0}=\Sigma\times \{0\}$} \\  \label{wave2}
  \dot{u}(0,x)&=0 \mbox{ on $\Sigma_{0}=\Sigma\times \{0\}$}
\end{align} 
where $f:M\rightarrow\mathbb{R}$ is a smooth function. Similarly the
retarded zero initial data inhomogeneous problem for the wave equation
on ${\overline{M}}$ is given by
\begin{align}\label{waveret}
  \square_{g}u&=f \mbox{ on $M$}\\
 u(T,x)&=0 \mbox{ on $\Sigma_{T}=\Sigma\times \{T\}$} \\  
  \dot{u}(T,x)&=0 \mbox{ on $\Sigma_{T}=\Sigma\times \{T\}$}
\end{align} 
 where $f:M\rightarrow\mathbb{R}$ is a smooth function.

 If the spacetime $(M,g_{ab})$ is globally hyperbolic the advanced and
 retarded zero initial data inhomogeneous problems are both well posed
 i.e. there exist unique solutions $u\in C^{\infty}(M)$ which depend
 continuously on $f$ \cite{bar}.  Moreover, in the globally hyperbolic
 case well-posedness is equivalent to the existence of unique advanced
 and retarded Green operators \cite{bar}. For convenience we recall
 the definition of these \cite{bar}.

\begin{Definition}\label{propagators}
 A linear map $E^{+}:D(M)\rightarrow C^{\infty}(M)$ satisfying
 \begin{itemize}
   \item $\square_{g}E^{+}=id_{D(M)}$
   \item $E^{+}\square_{g}|_{D(M)}=id_{D(M)}$
   \item $supp(E^{+}\psi)\subset J^{+}(supp(\psi))$ for all $\psi\in D(M)$
 \end{itemize}
 \noindent
 is called an advanced Green operator for $\square_{g}$. The retarded Green operators $E^{-}$ are defined in a similar fashion.
\end{Definition}
 
For a smooth globally hyperbolic spacetime the splitting theorem says
that the metric may be written in the form
\begin{equation}\label{ultra}
ds^{2}=+N^2  dt^{2}-{\gamma}_{ij}dx^{i}dx^{j}
\end{equation}
and this remains true even in low regularity \cite{clemens}. 
The aim of the paper is to define such advanced and retarded Green
operators for the metrics of the above form where the lapse function
$N$ and the induced Riemannian metric ${\gamma}_{ij}$ are no longer
smooth.

{\bf{Notation.}} We denote the derivative of a function $u$ with
respect $t$ by $u_{t}$ or $\partial_{t}u$ and $u_{i}$ or
$\partial_{i}u$ if it is with respect to the other
$x^i$-coordinates. The space of smooth functions of compact support
will be denoted by $D(M)$. A function $f$ on an open set $\cal{U}$ of
$\mathbb{R}^{n}$ is said to be Lipschitz if there is some constant $K$
such that for each pair of points $p,q\in {\cal{U}}$, $\arrowvert
f(p)-f(q)\arrowvert\le K\arrowvert p-q\arrowvert$, where $\arrowvert
p\arrowvert$ denotes the usual Euclidean distance. We denote by
$C^{k,1}$ those $C^{k-1}$ functions where the $k$-derivative is a
Lipschitz function.  A function $f$ on $M$ is said to be Lipschitz or
$C^{k,1}$ if $f$ is Lipschitz or $C^{k,1}$ in some coordinate chart.

In the analysis below we will be working with spaces such as
$L^2(\Sigma)$, $H^1(\Sigma)$ $H^{-1}(\Sigma)$ which are defined with
respect to a smooth background Riemannian metric $h_{ij}$ on $\Sigma$ with $\nu_h$
the corresponding volume form. We then define $L^2(\Sigma)$ to be
the space of real valued functions $g$ on $\Sigma$ such that
$\int_\Sigma g^2 \nu_h < \infty$ and we denote the associated inner
product by $(f,g)_{L^2}=\int_\Sigma fg\nu_h$. The space $H^k(\Sigma)$
are the real valued functions $g$ such that their first $k$
derivatives are in $L^{2}(\Sigma)$ and the space $H^{-k}(\Sigma)$ are
the bounded linear functionals on $H^k(\Sigma)$.

We also define the space $L^2(M,g)$ to be the space of real valued
functions $\phi$ on $M$ such that $\int_M \phi^2 \nu_g < \infty$ which
is defined with respect the volume form given by the metric
(\ref{ultra}) and $L^2(\Sigma_t,\gamma)$ to be the space of real
valued functions $\psi$ on $\Sigma_t=\{t\}\times \Sigma$ for $0\le
t<T$ such that $\int_{\Sigma_t} \psi^2 \nu_\gamma < \infty$ which is
defined with respect the volume form given by the Riemannian metric
$\gamma_{ij}$ that appears in (\ref{ultra}) . We denote in a
similar way the Sobolev space $H^k(M, g)$ and $H^k(\Sigma, \gamma)$.

We will also often think of a function ${\bf v}(t,x)$ as a map from
$[0,T]$ to a function ${\bf v}(t)(\cdot)$ in some Hilbert space
$X(\Sigma)$ given by ${\bf v}(t)(x)={\bf v}(t,x)$.  For example
$L^2(0, T; X(\Sigma))$ is the space of functions
\begin{eqnarray} \label{nested}
{\bf v}:[0,T] & \to & X(\Sigma)\\
t & \mapsto & {\bf v}(t)
\end{eqnarray} 
such that ${\bf v}(t) \in X(\Sigma)$ and
\begin{equation}
\left(\int_0^T \arrowvert\arrowvert{\bf v}(t)\arrowvert\arrowvert_{X(\Sigma)}^2 dt \right)^\frac{1}{2}< \infty
\end{equation}
When thinking of ${\bf v}$ in this way,
we will denote the time derivative by ${\bf \dot v}$.

We will also use the spaces $C^k([0, T]; X(\Sigma))$ which is the
space of functions
\begin{eqnarray} \label{nested}
{\bf v}:[0,T] & \to & X(\Sigma)\\
t & \mapsto & {\bf v}(t)
\end{eqnarray} 
such that ${\bf v}(t) \in X(\Sigma)$ and
\begin{equation}
\max_{j=0,...,k}\sup_{t\in[0,T]}\arrowvert\arrowvert \partial_{t^{j}} {\bf v}(t)\arrowvert\arrowvert_{X(\Sigma)}< \infty 
\end{equation}

where $\partial_{t^{l}} {\bf v}(t)$ is the $l$- derivative with
respect time of ${\bf v}(t)$.


\subsection{Generalised hyperbolicity}

To prove the existence of Green operators in the non-smooth setting we
will require that the spacetime satisfy the condition of
{\emph{generalised hyperbolicity}} which is essentially the requirement
that solutions to the wave equation are well-posed \cite{Clarke}. In
the low regularity setting the pointwise equation (\ref{wave}) may not
make sense and therefore a notion of a weak solution is required.

Taking into account the form of the metric (\ref{ultra}) and the
geometric conditions that appear in Theorem \ref{gws} (see below) the
wave operator may be written as
\begin{equation}
    \square_{g}u=\frac{\ddot{u}}{{\gamma}}-\frac{Lu}{{{\gamma}}}\label{weak11}
\end{equation}
where $-L$ is an elliptic operator in divergence form given by:

\begin{equation}
  -Lu=-({\gamma}^{ij}{{\gamma}}u_{j})_{i}.
  \end{equation}

We can associate with the operator $-L$ the following bilinear form given by:
\begin{equation}
  B[u,v; t]:=\int_{\Sigma}{\gamma}^{ij}(x,t){{\gamma}(x,t)} u_{i}(x)v_{j}(x) dx^{n}
\end{equation}

Using this bilinear form, we make the following definition of an
advanced weak solution.

\begin{Definition}\label{def}
We say a function:

$$u\in L^2([0,T]; H^1(\Sigma)), \mbox{ with } \dot{u}\in L^2([0,T]; L^{2}(\Sigma)), \ddot{u}\in L^2([0,T]; H^{-1}(\Sigma))$$
is an advanced weak solution of the zero initial data inhomogeneous
problem (\ref{wave}), (\ref{wave1}), (\ref{wave2}) provided that:

\begin{enumerate}
\label{weakcs}
  \item For each $v\in L^2([0,T]; H^1(\Sigma))$,
  \begin{equation}\label{weakf}
  \begin{split}
    \int^{T}_{0}<\ddot{u},v>dt+\int^{T}_{0}B[u,v;t]dt=(f,v)_{L^{2}(M,{g})}\end{split}
  \end{equation} where  $<\cdot,\cdot>$ denotes the dual pairing between the $H^{-1}(\Sigma)$ and $H^{1}(\Sigma)$ Sobolev spaces. 
  \item $u(0,x)=0, \dot{u}(0,x)=0$ \label{weak22}  
\end{enumerate}

\end{Definition}
The definition of the retarded weak solution is analogous.

\begin{rem}\label{bl}
  The regularity needed to make sense of Equation (\ref{weakf}) is that
$$
u\in L^2([0,T]; H^1(\Sigma)), \mbox{ with } , 
\ddot{u}\in L^2([0,T]; H^{-1}(\Sigma)).
$$ 
Therefore for every $ u\in L^2([0,T]; H^1(\Sigma)) \mbox{ with }
\ddot{u}\in L^2([0,T]; H^{-1}(\Sigma))$ we can define the bounded
linear functional $(\square_g u)[\cdot]$ on $L^2([0,T];H^{1}(\Sigma))$
given by

\begin{equation}
(\square_g u)[v]= \int^{T}_{0}<\ddot{u},v>dt+\int^{T}_{0}B[u,v;t]dt.
\end{equation}
for all $v\in L^2([0,T]; H^{1}(\Sigma))$ and therefore it make sense
to denote the pairing $\langle \square_g u, v\rangle_{L^2 H^{-1}, L^2  H^1}.$ 

\end{rem}

We say an advanced weak solution is regular if it satisfies a suitable
energy estimate. 
\begin{Definition} \label{regularc2} {(\bf Regular Weak Solution)} 

We say an advanced weak solution is {\bf regular} if ${u}$ satisfies
the energy estimate
  \begin{equation}
\begin{split}
 \displaystyle ||u(t,\cdot)||_{L^2([0,T]; H^{1}(\Sigma))}+||\dot{u}(t,\cdot)||_{L^2([0,T]; L^{2}(\Sigma))}+||\ddot{u}(t,\cdot)||_{L^2([0,T]; H^{-1}(\Sigma))}\\
      \le C||f||_{L^{2}(M,g)}
      \end{split}
\end{equation}
\end{Definition}

We will also require the weak solutions to satisfy certain support
properties. In the smooth case the condition of global hyperbolicity
is such that a solution $u$ of the inhomogeneous problem satisfies
$supp(u)\subset J^{+}(supp (f) )\cup J^{-}(supp (f) )$. We use this
observation and the definition of the support of a distribution $u$
i.e. $supp(u)=\{x\in M|$ for all neighbourhoods $U$ of $x$ there exist
$\varphi \in D(M)$ with $supp(\varphi)\subset U$ and $u[\varphi]\neq
0\}$ to make the following definition:

\begin{Definition} \label{regularc3} {(\bf Causally supported Weak Solution)} 
We say a weak solution is {\bf causally supported to the future } if
  ${u}$ satisfies the support condition
  \begin{equation}
supp(u)\subset J^{+}(supp (f) )
\end{equation}
in the sense of distributions. The definition of causally supported to
the past is analogous. We say a weak solution is {\bf causally supported }
if it is causally supported to both the past and future.
\end{Definition}

We are now in a position to extend the notion of generalised
hyperbolicity to the case where the wave equation only admits weak solutions.

\begin{Definition} \label{genhyp} {(\bf Generalised hyperbolicity)} 
We say a spacetime $M$ satisfies the advanced generalised
  hyperbolicity condition if the wave equation is well-posed in the
  following sense: There exists a unique regular weak solution $u$ of
  $\square_{g}u=f$ in the sense of definition \ref{def} and definition
  \ref{regularc2} with initial conditions
\begin{enumerate}
   \item  $u(0,\cdot)=u|_{\Sigma_{0}}=0$  
  \item $\dot{u}(0,\cdot)=\frac{\partial u}{\partial t}|_{\Sigma_{0}}=0$
  \end{enumerate}
  that is causally supported to the future.  With the obvious
  modifications we define the condition of retarded generalised
  hyperbolicity. If a spacetime satisfies the advanced and retarded
  condition of generalised hyperbolicity we say the spacetime
  satisfies the condition of generalised hyperbolicity.
\end{Definition}

\begin{rem}
  In the case of a smooth metric, global
  hyperbolicity is sufficient to guarantee that the condition of
  generalised hyperbolicity in the above sense is also satisfied.
\end{rem}

The following series of theorems give sufficient conditions to ensure
such that a low regularity spacetime $(M, g) $ satisfies the condition
of generalised hyperbolicity. For simplicity, we only deal with the
condition of advanced generalised hyperbolicity.

\begin{Theorem}\label{gws}
  Let the metric $g_{ab}$ given by Equation \ref{ultra} on $M$ satisfy
  the following conditions

\begin{enumerate}
  \item[1.] $\gamma^{ij}\in C^{1}([0,T], L^{\infty}(\Sigma))$.
  \item[2.] The scalar coefficient of the volume form given by
    $\sqrt{\gamma}$ for the induced metric $\gamma_{ij}$ is bounded
    from below by a positive real number, i.e., $|\sqrt{\gamma}|>\eta$
    for some $\eta\in \mathbb{R}^{+}$
  \item[3.] The lapse function $N$ can be chosen as $N=\sqrt{\gamma}$ .
  \item[4.] There exist a constant $\theta>0$ such that
  $$\sum^{n}_{i,j=1}\gamma^{ij}{\gamma}\xi_{i}\xi_{j}\ge\theta|\xi|^{2}$$
  for all $(t,x)\in M, \xi\in\mathbb{R}^{n}$.
  \end{enumerate}
  then for every $f\in L^{2}(M,g)$ there are advanced and retarded
  weak regular solutions
\end{Theorem}

\begin{rem}
  Condition 3 is chosen for simplicity. However the condition can
  be weakened to require only that it is a bounded function with a
  positive lower bound, i.e.
\begin{equation*}
 \emph{$3'.$} \mbox{ \emph{The lapse function} } N  \mbox{\emph{ is} } C^{1}([0,T], L^{\infty}(\Sigma)) \mbox{ \emph{and} }  |N|>\omega \mbox{ \emph{for some} } \omega\in \mathbb{R}^{+}
 \end{equation*}
This modification is at the expense of adding additional terms and
to avoid undue complications in the formulae we do not pursue this
here.
\end{rem}

\noindent
{\emph{Proof}} The main idea is to use Galerkin's method. We sketch
the proof below for a complete proof see \cite{evans, hunter, cs}.

\begin{itemize}
\item Introduce an orthogonal basis in $H^1(\Sigma)$
  $\{w_k\}_{k=1}^\infty$ which is also an orthonormal basis in
  $L^2(\Sigma)$.
\item Insert the approximate solution $u^{(m)}(t,x)=\sum_{k=1}^m d^{(m)}_k(t) w_k(x)$ \\
  into the projected wave equation $( u_{tt}^{(m)},
  w_k)_{L^{2}(\Sigma)}+B[u^{(m)},w_k]=({\gamma}
  f,w_k)_{L^{2}(\Sigma)}$ for all $k=1,...,m$. This gives a first
  order system of ODEs for $d^{(m)}_k(t)$ with zero initial data on
  $t=0$.
\item Prove that the $u^{(m)}$ satisfy the energy estimate uniformly
  in $m$
\item Apply Banach-Alouglu theorem to extract a subsequence which is a
  solution to the weak formulation
\item Use lower semi-continuity of the norm and energy estimates to
  prove uniqueness, stability and regularity of the weak solution.
\end{itemize}

\begin{Theorem}\label{gh}
  Let the metric $g_{ab}$ be a $C^{1,1}$ globally hyperbolic spacetime
  metric on $M$ and the source function $f$ be an element of
  $H^{1}(M,g)$ then $(M, g)$ satisfies the condition of generalised
  hyperbolicity.
\end{Theorem}
\noindent
{\emph{Proof.}} 
The results of \cite{evans, hunter, cs} establish well-posedness. The
only thing that remains to be shown is that the weak solutions have
the correct support. To do so we follow \cite{hawking} and obtain an
energy inequality that determines the support of the solution. 
      
Consider a weak solution $u$ with energy-momentum tensor
\begin{equation}
       T^{ab}[u] = \left(g^{ac}g^{bd}-\frac{1}{2}g^{ab}g^{cd}\right)\frac{\partial u}{\partial x^{c}}\frac{\partial u}{\partial x^{d}}\label{energytensor}
\end{equation}
If we contract this with a Lipschitz timelike vector field, $\Upsilon^{a}$
and use the fact that $J^{-}_{\Sigma_t}(K)\cap J^{+}(\Sigma_0)$  is
compact by causal theory for globally hyperbolic $C^{1,1}$ metrics
\cite{cru,c11, clemens} (where $J^{-}_{\Sigma_t}(K)$ denotes the
causal past of a compact region $K\subset \Sigma_t$), then provided we have sufficient regularity that one can apply the divergence theorem
we obtain
\begin{equation}\label{stokess}
  \int_{J^{-}_{\Sigma_t}(K)\cap J^{+}(\Sigma_0)}\mbox{div}\left(T^{ab} \Upsilon_{a}\right)\nu_{g}=\int_{\partial(J^{-}_{\Sigma_t}(K)\cap J^{+}(\Sigma_0))}T^{ab}\Upsilon_{a}n_{b}\nu_{\gamma}
\end{equation}
In order for the above equation to be well-defined we require that
$\mbox{div}(T^{ab}\Upsilon_a)$ should be integrable with respect the volume
form $\nu_g$, and this in turn requires that the weak solutions are
sufficiently regular. In fact it is enough that the weak solutions
have two derivatives in $L^{2}(M,g)$ if the metric and the timelike
vector field are in the space $C^{0,1}$ \cite{crusdiv} and this is ensured by 
the additional regularity of the metric $g \in C^{1,1}$ and source $f \in H^1 (M,g)$ required in the conditions of Theorem \ref{gh} compared to Theorem \ref{gws}.
Moreover, such an improvement in regularity also guarantees that $u$
belongs to the space $C^0([0,T], H^{1}(\Sigma))\cap C^1([0,T],
L^{2}(\Sigma))$. See \cite{evans} for a proof of these results.

The left hand side of \ref{stokess} takes the explicit form:
\begin{equation}\label{left0}
    \int_{J^{-}_{\Sigma_t}(K)\cap J^{+}(\Sigma_0)}\left(\left(g^{ab}\frac{\partial u}{\partial x^{b}}\Upsilon_{a}\right)\left[\square_g u-u\right]+T^{ab}\nabla_{b}\Upsilon_{a}\right)\nu_g
\end{equation} 
where $\nabla_{b}\Upsilon_{a}$ denotes the covariant derivative of $\Upsilon_{a}$ while the right hand side takes the form:
\begin{equation}
\left(\int_{K}+\int_{\Sigma_{0}}\right)T^{ab}\Upsilon_{a}n_{b}\nu_{\gamma}+\int_{\cal{H}}T^{ab}\Upsilon_{a}n_{b}\nu_{h}.
\end{equation}
where ${\cal{H}}=\partial(J^{-}_{\Sigma_t}(K)\cap
J^{+}(\Sigma_0))\backslash (\Sigma_0\cup K)$, $\nu_h$ is the induced
volume form on ${\cal{H}}$ and $n^{b}$ denotes outward pointing normal
vectors on $K$, $\Sigma_0$ and $\cal{H}$.

Moreover, in the globally hyperbolic $C^{1,1}$ case we have that
${\cal{H}}$ is a null hypersurface ruled by null geodesics \cite{c11}
and by the dominant energy condition \cite{hawking} we have that
\begin{equation}
\int_{\cal{H}}T^{ab}\Upsilon_{a}n_{b}\nu_{h}\ge 0
\end{equation}
For the integral over $K$ we define the energy integral:
\begin{equation}
 E_K(t)=\int_{K}T^{ab}\Upsilon_{a}n_{b}\nu_{\gamma}
\end{equation}
which is equivalent to a Sobolev type norm on $K$ 
\begin{equation}
\tilde{\Arrowvert} u\Arrowvert^{1}_{K}=\left[\int_{K}\left(\left(\frac{\partial u}{\partial t}\right)^{2}+\sum^{3}_{i=1}\left(\frac{\partial u}{\partial x^{i}}\right)^{2}\right)\nu_{\gamma}\right]^{\frac{1}{2}}
\end{equation}
since 
\begin{equation}\label{wilson11}
C_{1}E_K(t)\le(\tilde{\Arrowvert} u\Arrowvert^{1}_{K})^{2}\le C_{2} E_K(t)
\end{equation}
for constants $C_{1},C_{2}\ge0$ \cite{conical}. In a similar way we may relate energy integral $E_{\Sigma_0}$ and Sobolev norms $\tilde{\Arrowvert}
u\Arrowvert^{1}_{\Sigma_0}$ on $\Sigma_0$.

We now use the fact that 
\begin{equation}\label{wilson22}
  \Arrowvert\phi\Arrowvert_{{{H^{1}(J^{-}_{\Sigma_{t'}}(K)\cap J^{+}(\Sigma_0),g)}}}\le C \left(\int_{0}^{t}\left(\tilde{\Arrowvert} \phi\Arrowvert^{1}_{K}\right)^{2}dt'\right)^{\frac{1}{2}}
\end{equation}
where $C$ is a constant that depends on $g_{ab}$ and the interval $[0,t]$ \cite{conical}.
Estimating all the terms in (\ref{stokess}) by using the  Cauchy Schwarz inequality
together with the regularity of the metric and solutions gives the inequality:
  
\begin{equation}\label{preeee}
    E_K(t)\le E_{\Sigma_0}(0) + k_{0} (\Arrowvert \square_g u\Arrowvert_{L^{2}(J_{\Sigma_t}^{-}(K)\cap J^{+}(\Sigma_0),{g})})^{2} + k_{1}(\Arrowvert u\Arrowvert_{H^{1}(J_{\Sigma_t}^{-}(K)\cap J^{+}(\Sigma_0),{g})})^{2}
\end{equation}
where $k_{0},k_{1}$ are positive constants that depend on the metric
$g_{ab}$, the vector field $\Upsilon_{a}$ and the covariant derivative
$\nabla_{b}\Upsilon_{a}$.
    
Now rewriting (\ref{preeee}) using (\ref{wilson11}) and
(\ref{wilson22}) we find
  
\begin{equation}
   E_K(t)\le E_{\Sigma_{0}}(0) + k_{0}(\Arrowvert \square_g u\Arrowvert_{L^{2}(J_{\Sigma_t}^{-}(K)\cap J^{+}(\Sigma_0),\nu_{g})})^{2} +  k_{2}\int^{t}_{0}E(t')dt'
  \end{equation} 
  
Using Gronwall's inequality we obtain
\begin{equation}
     E_K(t) \le K_{4} \left(E_{\Sigma_0}(0)+ (\Arrowvert \square_g u\Arrowvert_{L^{2}(J_{\Sigma_t}^{-}(K)\cap J^{+}(\Sigma_0),{g})})^{2}\right)\label{ee1}
              \text{ for all } t\le T
\end{equation}
where $K_{4}$ is a positive constant that depends on the chosen finite
time $t$, the metric $g_{ab}$, the vector field $\Upsilon_{a}$ and the
covariant derivative $\nabla_{b}\Upsilon_{a}$.
 
In term of the Sobolev norms we obtain the expression:
\begin{equation}
( \tilde{\|} u\|^{1}_{K})^{2} \le A \left (( \tilde{\|} u\|^{1}_{\Sigma_{0}})^{2} + (\Arrowvert \square_g u\Arrowvert_{L^{2}(J_{\Sigma_t}^{-}(K)\cap J^{+}(\Sigma_0),{g})})^{2}\right)
  \end{equation} 
for some constant $A$.
   
By definition we have that the advanced weak solution have vanishing
initial data, therefore $\tilde{\|} u\|^{1}_{\Sigma_{0}}=0$ and we
obtain the energy estimate
\begin{equation}\label{supp}
( \tilde{\|} u\|^{1}_{K})^{2} \le A  (\Arrowvert \square_g u\Arrowvert_{L^{2}(J_{\Sigma_t}^{-}(K)\cap J^{+}(\Sigma_0),{g})})^{2}
  \end{equation} 
This establishes the fact that an advanced solution $u$ of the zero
  initial data inhomogenous problem $\square_g u=f$ on
  $K\subset\Sigma_t$ is determined by the value of the source $f$ in
  the region $J^{-}_{\Sigma_t}(K)\cap J^{+}(\Sigma_0)$.

  Notice that if $p$ does not belongs to a neighbourhood of $J^{+}(supp
  (f))$ then $J^-(p)$ does not intersect $supp(f)$ and therefore we
  can construct a compact set $K\subset\Sigma_{t'}$ that contains $p$
  for some $t'$ such that $J^{-}_{\Sigma_t'}(K)\cap J^{+}(\Sigma_0)$
  does not intersect $supp(f)$. This implies that the right hand side
  of (\ref{supp}) vanishes and $u=0$ on $K$. Then, taking a possible
  smaller region $K'\subset K$ and a small time interval $\epsilon$ we
  can form a neighbourhood $U=\epsilon\times K'$ that contains $p$ and
  where $u$ vanishes. For example, it is sufficient to take points
  $q,r$ such that $q\in J^+(K)$ and $r\in J^-(K)$ such that $q,r$ does
  not intersect $J^{+}(supp (f))$ and consider the region $J^-(q)\cap
  J^+(r)$. In this region we have that $u=0$ and this implies that $p$
  is not in the support of $u$ which gives the desired result.

\begin{rem}\label{rem}
  It may be possible to relax the regularity conditions to those of
  Theorem \ref{gws} by analysing the causal structure of spacetimes
  with low regularity as in \cite{cru, c11, clemens} and using results
  on propagation of singularities for differential equations with
  rough coefficients as in \cite{rough}.
\end{rem}


\section{Green operators in the non-smooth setting}

In this section we define the notion of advanced and retarded Green
operators in a weak sense. Notice that the Definition
\ref{propagators} can not be used in a setting with finite
differentiability because weak solutions are in general not smooth so
that one cannot expect $E^+(f) \in C^\infty(M)$. We therefore propose
the following definitions suitable for settings with finite
differentiability.
\begin{Definition} \label{Green}
 A bounded linear map $E^{+}:L^{2}(M,g)\rightarrow L^2([0,T]; H^1(\Sigma))$ satisfying
 \begin{itemize}
   \item $\langle\square_{g}E^{+}(f), v\rangle_{L^2 H^{-1}, L^2 H^1}=(f,v)_{L^2(M,g)}$ for all $f\in D(M), v\in L^2([0,T]; H^1(\Sigma))$
   \item $\langle E^{+}\square_{g}(f),v\rangle_{L^2 H^{-1}, L^2 H^1}=(f,v)_{L^2(M,g)}$ for all $f\in D(M), v\in L^2([0,T]; H^1(\Sigma))$
   \item $supp(E^{+}(f))\subset J^{+}(supp(f))$ for all $f\in D(M)$
 \end{itemize}
 \noindent
 is an advanced weak Green operator for $\square_{g}$. A retarded weak
 Green operator $E^{+}$  is defined similarly.
\end{Definition}

\begin{rem}\label{inv1}
  If $g_{ab}$ is in $C^{0,1}$ and for all $f\in D(M)$ we have  $E^{\pm}(f)$ in $H^{2}(M,g)$ then the first property can be restated as

\begin{equation}
(\square_{g}E^{\pm}f, v)_{L^2(M,g)}=(f,v)_{L^2(M,g)}
\end{equation}
 for all $f\in D(M), v\in L^2([0,T]; H^1(\Sigma))$.
\end{rem}
\begin{rem}\label{inv2}
  If for all $f\in D(M)$ we have $E^{\pm}(\square_g f)$ in $L^{2}(M,g)$
  then the second property can be restated as

\begin{equation}
(E^{\pm}\square_{g}f, v)_{L^2(M,g)}=(f,v)_{L^2(M,g)}
\end{equation}
 for all $f\in D(M), v\in L^2([0,T]; H^1(\Sigma))$
\end{rem}

We now show that the notion of generalised hyperbolicity is sufficient
to establish the existence of weak Green operators.

 \begin{Theorem}
   Let $(M, g)$ be a globally hyperbolic spacetime with $g_{ab} \in
   C^{1,1}$. Then on $M$ there are unique advanced and retarded weak
   Green operators for $\square_g$.
 \end{Theorem}
 
\noindent
 {\emph{Proof.}}  We will only show the existence of the advanced weak
 Green operator. The existence of the retarded weak Green operator
 follows from time reversal.

 Define the linear map $E ^{+}:L^{2}(M,g) \rightarrow L^2([0,T];
 H^1(\Sigma))$ which send a source function $f$ to the advanced
 regular weak solution $u^{+}$ of the advanced zero initial data
 inhomogeneous problem with source $f$ . That such $ u^{+}$ exist and
 is unique is a consequence of generalised hyperbolicity which follows
 from Theorem \ref{gh}.

We first show that this map is well defined and continuous.

Let $f=g\in L^{2}(M,g) $, then by definition we have that
$E^{+}(f)=u^{+}$ and $E^{+}(g)=v^{+}$ are regular weak solutions of
the zero initial data inhomogeneous problem with source function $f=g$
causally supported to the future. By the condition of generalised
hyperbolicity which guarantees uniqueness of solutions, we have
$u^{+}=v^{+}$ and therefore $E^{+}(f)=E^{+}(g)$. Therefore the map is
well defined.

That the map is continuous follows from the regularity of solutions. We have 

\begin{equation}
\arrowvert\arrowvert E^{+}(f)\arrowvert\arrowvert_{L^2([0,T];H^{1}(\Sigma))}=\arrowvert\arrowvert u\arrowvert\arrowvert_{L^2([0,T]; H^1(\Sigma))}\le\arrowvert\arrowvert f\arrowvert\arrowvert_{L^{2}(M,g)}
\end{equation}

We  next show that $E^{+}$ is a suitable right inverse in the sense of Definition \ref{Green}.  First notice
that the regularity of the weak solutions given by the condition of
generalised hyperbolicity allows us to define for every advanced weak
solution $u^{+}$ the linear map $\square_g u^{+} [\cdot]$.  See remark
\ref{bl}.

Therefore given $E^{+}(f)=u^{+}$ , we have 
\begin{align}
\square_g E^{+}(f)[v]=: \square_g u^{+}[v]= \begin{split}
    \int^{T}_{0}<\ddot{u}^{+},v>dt+\int^{T}_{0}B[u^{+},v; t]dt=(f,v)_{L^{2}(M,g)}\end{split}
\end{align}
which gives
\begin{align}
\langle\square_{g}E^{+}f, v\rangle_{L^2 H^{-1}, L^2 H^1}=(f, v)_{L^2(M,g)}
\end{align}
for all $f\in D(M), v\in L^2([0,T];H^{1}(\Sigma))$

Now we prove a similar equality for $E^+ \square_g f [v]$ to show that
it is a left inverse. First we notice that by definition we have that
$u^{+}=E^{+}(\square_g f)$ is an advanced weak solution of the
inhomogeneous problem with source $\square_g f$ and therefore we have
that
\begin{equation}
\int^{T}_{0}<\ddot{u}^{+},v>dt+\int^{T}_{0}B[u^{+},v;t]dt=(\square f,v)_{L^{2}(M,{g})}
\end{equation}

At the same time using integration by parts we have that 
\begin{equation}
\int^{T}_{0}<\ddot{f},v>dt+\int^{T}_{0}B[f,v;t]dt=(\square f,v)_{L^{2}(M,g)}
\end{equation}
so $f$ is also a weak solution.  Moreover, using the smoothness of $f$
and the assumed regularity of the metric we can obtain an energy
estimate which proves that $f$ is a regular weak solution (see
\cite{evans}). Then $u^{+}$ and $f$ are weak regular solutions to the
zero initial data inhomogeneous problem with the same source. So by uniqueness of the solutions we
have that $u^{+}=f$ which guarantees that
\begin{align}
\langle E^{+}\square_{g}f,v\rangle_{L^2 H^{-1}, L^2 H^1}=(f,v)_{L^2(M,g)}
\end{align}
 for all $f\in D(M), v\in L^2([0,T];H^{1}(\Sigma))$

 Finally to analyse the support of $E^{+}(f)$ we use that the weak solutions are
 causally supported. We have that $supp(E^+ (f))=supp(u^{+})\subset
 J^{+}(supp (f))$ for all $f\in D(M)$ which gives the desired result.

\begin{rem}
  It is pointed out in remark 3.4.9 in \cite{bar} that Green operators
  on compact spacetimes with smooth metrics cannot exist as that
  would imply that $\square_g:D(M)\rightarrow D(M)$ is injective. The
  proof uses the fact that if $\square_g \phi=0$ and Green operators exist,
  then one has $\phi=E^{\pm}\square_g\phi=E^{\pm} 0=0$ which shows
  injectivity. However, any constant function is in the kernel of
  $\square_g$.
  In our case, if one tries to repeat such a result one has the
  following equalities $(\phi,v)=(E^{\pm}\square_{g}\phi,
  v)_{L^2(M,g)}=(E^{\pm}0, v)_{L^2(M,g)}=(0,v)_{L^2(M,g)}=0$.
By the density of $ L^2([0,T]; H^1(\Sigma))$ in $L^{2}(M,g)$ we can  conclude that $\phi=0$.
which recovers the non-existence result in our setting also and explains the choice $M=(0,T) \times \Sigma$.
\end{rem}

We next show that the formal adjoint of $E^{+}$ is $E^{-}$.

\begin{Theorem}\label{greenadj}
Given $\chi\in D(M)$ and $\varphi\in D(M)$ we have that
\begin{equation}
\int_M E^+(\chi) \varphi \nu_g =\int_M \chi E^-(\varphi) \nu_g 
\end{equation}
\end{Theorem}

In order to show this result we need the following Lemma

\begin{Lemma}\label{selfadj}
Given $\chi, \varphi\in H^{2}(M,g)$ with $\chi(0,x)=\chi_t(0,x)=0$ and $\varphi(T,x)=\varphi_t(T,x)=0$  we have that
\begin{equation}
\int_M \square_g \chi \varphi \nu_g =\int_M \chi \square_g\varphi \nu_g 
\end{equation}
\end{Lemma}
\noindent
{\emph{Proof.}}  The proof follows straightforward from using
integration by parts twice and using the boundary conditions given by
the hypothesis.

\noindent
{\emph{Proof of Theorem \ref{greenadj}.}}

First, notice that because the metric has regularity $C^{1,1}$ and
$\chi,\varphi\in D(M)$ we have that $ E^+(\chi), E^-(\varphi)\in
H^{2}(M,g)$. \cite{evans}.

Hence,

\begin{align}
\int_M E^+(\chi) \varphi \nu_g &=(E^+(\chi),  \varphi )_{L^{2}(M,g)}\\
&=(E^+(\chi),\square_g E^-(\varphi) )_{L^{2}(M,g)}\\
&=(\square_g E^+(\chi), E^-(\varphi) )_{L^{2}(M,g)}\\
&=(\chi, E^-(\varphi) )_{L^{2}(M,g)}\\
&=\int_M \chi E^-(\varphi) \nu_g 
\end{align}
where the second and fourth equality follows from remark \ref{inv1}
while the third inequality used the initial data given by the advanced
and retarded zero initial data inhomogeneous problem and Lemma
\ref{selfadj}.  

In the smooth case it is often convenient to consider the Green operators as
bi-distributions on the product space. We show that this result extends to the non-smooth setting.
 \begin{Theorem}\label{bi}
The map ${\bf{E}}^+:D(M\times M) \rightarrow \mathbb{R}$ given by 

\begin{equation}
{\bf{E}}^+ (f,g)=(E^+(f),g)_{L^2 (M,g)}
\end{equation}
 is a distribution on $M \times M$. 
 \end{Theorem}
\noindent
{\emph{Proof of Theorem \ref{bi}.}}
%
%
%
%
%
%
%
%
%

Given $f\otimes g\in D(M)\otimes D(M)$ we have 
\begin{align}
{\bf{E}}^+ (f, g)&=(E^+(f),g)_{L^2 (M,g)}\\
&\le||E^+(f)||_{L^2 (M,g)}||g||_{L^2 (M,g)}\\
&\le||E^+(f)||_{L^2([0,T],H^{1}(\Sigma)) }||g||_{L^2 (M,g)}\\
&\le C_1||f||_{L^2(M,g)}||g||_{L^2 (M,g)}\\ 
&\le C_1|||f\otimes g||_{L^2 (M,g)}\\ 
&\le C_2 ||f\otimes g||_{C^0 (M)}\\ 
\end{align}
for some constants $C_1, C_2$ and where we have used Holder's
inequality and that $E^{+}(f)$ is a regular solution. Now taking into
account that $D(M)\otimes D(M)$ is dense in $D(M\times M)$ we can
extend ${\bf{E}}^+ (f, g)$ to a unique continuous linear mapping in
$D(M\times M)$. This proves the claim.

\begin{rem}
See \cite{bar} for a proof of this theorem in the smooth setting.
\end{rem}


\section{Galerkin type approximation of $E^{+}$}

In this section we will provide a construction of the operator $E^{+}$
and give a formula for the kernel. To construct the map $E^{+}$ we will build a sequence of operators
$E_m^{+}$ that converge in the weak operator topology to $E^{+}$. We
will construct the operators $E_m^{+}$ using Galerkin's method
following the methods of \cite{evans, hunter, cs}.

First consider a complete orthogonal basis $\{w_j(x)\}_{j \in J}$ in
$H^1(\Sigma)$ which is also orthonormal in $L^2(\Sigma)$. Then
inserting the $m$-approximate solution $u^{(m)^{+}}(t,x)=\sum_{k=1}^m
d^{(m)}_k(t) w_k(x)$ into the equation \ref{weakf} we find that the
functions $d^{(m)}_k$ for $k=1,...,m$ satisfies the system of second
order ODEs
\begin{equation}
\ddot{d}^{k}_{m}(t)+\sum^{m}_{l=1}e^{kl}(t)d^{l}_{m}(t)=f^{k}(t)
\end{equation}
where $e^{kl}(t)=B[w_k,w_l;t]$ are $C^1$ and symmetric and $f^{k}(t)=\int_\Sigma f(t,y)w_j(y){\gamma}(y,s) dy$. (See Theorem \ref{gws})

We may find a Green matrix $F^{kj}_m(t,s)$ \cite{greenmatrix} for the initial
value problem for this system of second order ODEs and write the solution in the form
\begin{equation}
d^{k}_m(t)=\int_0^T \sum_jF^{kj}_m(t,s)f^j(s)ds 
\end{equation}
So that 
\begin{equation}
u^m(t,x)=\int_0^T\sum_{j,k}F^{kj}_m(t,s)f^j(s)w_k(x)ds
\end{equation}
Recalling that $f^j(s)=\int_\Sigma f(s,y)w_j(y){\gamma}(y,s)dy$ we have
\begin{equation}
u^m(t,x)=\int_0^T\int_\Sigma\sum_{j,k}F^{kj}_m(t,s)w_k(x)w_j(y)f(s,y){\gamma}(y,s)dyds
\end{equation}
We may now define the $m$-approximate advanced Green operator, $E_m^{+}$ on
$L^{2}(M,g)$ by
\begin{equation} \label{greenG}
E_m^{+}(f):=\int_0^T\int_\Sigma\sum_{j,k}F^{kj}_m(t,s)w_k(x)w_j(y)f(s,y){\gamma}(y,s)dyds
\end{equation}
where $F^{kj}_m(t,s)$ is the Green matrix for the (regular) ODE problem.

The next step is to show that $\{u^m(t,x)\}_m$ is bounded uniformly in
$L^2([0,T];H^{1}(\Sigma))$. This is a consequence of the energy
estimates. See \cite{evans} for a derivation of the estimates and the
uniform bound.  Applying the Banach-Alaoglu theorem to the bounded
sequence we can obtain a subsequence $\{u^{m_l}(t,x)\}_l$ that converges weakly in
$L^2([0,T];H^{1}(\Sigma))$. i.e. there is a subsequence
$\{E_{m_l}^+(f)=u^{m_l}\}_l$ that converges to a linear bounded
operator $E^+$ on $L^2([0,T];H^{1}(\Sigma))$ and its value on $v$
coincides with $u^+[v]$ for all $v\in L^2([0,T]; H^1(\Sigma))$.

We may therefore write
\begin{align}
u^{+}[v]&=E^{+}(f)[v]\\
&=\lim_{l\rightarrow\infty}\int_{M}E_{m_l}^+(f)(x,t) v(t,x){\gamma}(x,t) dtdx\\
&=\lim_{l\rightarrow\infty}\int_{M}\int_M\sum_{j,k}F^{kj}_{m_l}(t,s)w_k(x)w_j(y) \hfill f(s,y)v(t,x)\hfill {\gamma}(y,s){\gamma}(x,t) dyds dtdx
\end{align}
where the first identity is the weak convergence of the
$m$-approximate solutions $E_{m}^+(f) (x,t)=u^{m}(x,t)$ to the weak
solution $u^{+}(x,t)$ in $L^2([0,T]; H^1(\Sigma))$ and the second
identity follows from equation \ref{greenG}. We have therefore shown
that we may write the kernel of the advanced Green operator as
\begin{equation}
G^+(t, x; s,y)= \lim_{l\rightarrow\infty}\sum_{j,k}F^{kj}_{m_l}(t,s)w_k(x)w_j(y){\gamma}(y,s){\gamma}(x,t).
\end{equation}
Note that in terms of the above representation the fact that the
formal adjoint of $E^{+}$ is $E^{-}$ is reflected in the property
$F(s,t)=F^T(t,s)$ for the Green matrix of a symmetric self-adjoint
system of 2nd order ODEs \cite{greenmatrix}.

\ack

YSS would like to acknowledge  the financial support given by the Riemann fellowship. \\ JAV would like to acknowledge financial support by the UK STFC and the Austrian Science Fund FWF.



\section*{References}
\begin{thebibliography}{9}

\bibitem{bar} C. B{\"a}r, N. Ginoux, F. Pf{\"a}ffle {\emph{Wave Equations on Lorentzian Manifolds and Quantization}}  ESI  (2007).

\bibitem{fp} J. Derezi{\'n}ski, D. Siemssen ArXiv.1608.06441 (2016) 

\bibitem{Clarke} C J S Clarke {\emph{Class. Quantum Grav.}} {\bf{15}} 975 (1998)


\bibitem{hawking} S. W. Hawking and G. F. R. Ellis, {\emph{The large scale structure of space-time}},  Cambridge University Press, (1974)

\bibitem{evans} L. C. Evans {\it{Partial Differential Equations}}  American Mathematical Society, (2002).


\bibitem{hunter} J. Hunter \emph{Lecture Notes} \url{https://www.math.ucdavis.edu/~hunter/pdes/pde_notes.pdf}

\bibitem{cs}  {{Y. Sanchez Sanchez}}, J.A Vickers {\it {Class. Quant. Grav.}}  {\bf{33}} (2016)


\bibitem{cru}P. T. Chru\'sciel, J. D. E. Grant  {\emph{Class. Quantum Grav.}} {\bf{29}} 145001 (2012)


\bibitem{c11} M. Kunzinger, R. Steinbauer, M. Stojkovic, J. Vickers  \emph{Gen. Relativ. Gravit.} {\bf{46}} 1738 (2014)


\bibitem{clemens} C. S\" amann {\emph{Annal. Henri. Poincare}} {\bf{17}} 6 (2016)


\bibitem{conical} J. P. Wilson  {\emph{Class. Quantum Grav.}} {\bf{17}} 3199 (2000)

\bibitem{crusdiv} P. T. Chru\'sciel  \emph{Journal of Fixed Point Theory and Applications} {\bf 4}, 325 (2013)
\bibitem{rough} H. Smith \emph{Anal. PDE} {\bf{7}} 1137 (2014)

\bibitem{greenmatrix} T. \c{C}a\u{g}ri \c{S}i\c{s}man and B. Tekin  \emph{Journ. Phys. A} {\bf{43}} 12 (2010) 







\end{thebibliography}

\end{document}